\documentclass[twocolumn,epjc3]{svjour3}
\usepackage{graphicx}
\usepackage{grffile}

\usepackage[numbers]{natbib}
\usepackage[breaklinks=true]{hyperref}
\hypersetup{
    colorlinks=true,
    linkcolor=blue,
    citecolor=blue,
    filecolor=magenta,      
    urlcolor=cyan,
    pdftitle={},
    }
\journalname{Eur. Phys. J. A}

\begin{document}

\title{AGATA: Advancements in Software Developments}

\author {
O.~St\'ezowski\thanksref{IP2I, e1}
\and 
J.~Dudouet\thanksref{IP2I}
\and 
A.~Goasduff\thanksref{LNL}
\and 
A.~Korichi\thanksref{ijclab} 
\and 
Y.~Aubert\thanksref{ijclab,e2}
\and
M.~Balogh\thanksref{LNL}
\and
G.~Baulieu\thanksref{IP2I} 
\and 
D.~Bazzacco\thanksref{padova}
\and
S.~Brambilla\thanksref{milano} 
\and
D.~Brugnara\thanksref{LNL}
\and
 N.~Dosme\thanksref{ijclab}
\and 
S.~Elloumi\thanksref{ijclab}
\and 
P.~Gauron\thanksref{ijclab}
\and 
X.~Grave\thanksref{ijclab,e3}
\and 
J.~Jacob\thanksref{ijclab}
\and 
V.~Lafage\thanksref{ijclab}
\and
A.~Lemasson\thanksref{GANIL}
\and 
 E.~Legay\thanksref{ijclab}
\and 
P.~Le Jeannic\thanksref{ijclab}
\and 
J.~Ljungvall\thanksref{ijclab}
\and 
A.~Matta\thanksref{lpccaen}
\and 
R.~Molina\thanksref{ijclab}
\and 
G.~Philippon\thanksref{ijclab}
\and
M.~Sedlak\thanksref{LNL}
\and
M.~Taurigna-Quere\thanksref{ijclab}
\and
N.~Toniolo\thanksref{LNL}
}
\thankstext{e1}{e-mail: o.stezowski@ip2i.in2p3.fr}
\thankstext{e2}{Deceased}
\thankstext{e3}{The author is on leave from IJCLab-CNRS}

\institute{Universit\'e de Lyon 1, CNRS/IN2P3, UMR5822, IP2I, F-69622 Villeurbanne Cedex, France \label{IP2I}
\and
Istituto Nazionale di Fisica Nucleare, Laboratori Nazionali di Legnaro, Legnaro, Italy \label{LNL}
\and
Istituto Nazionale di Fisica Nucleare, Sezione di Milano, Milano, Italy \label{milano}
\and
Istituto Nazionale di Fisica Nucleare, Sezione di Padova, Padova, Italy \label{padova}
\and
IJCLab, Universit\'e Paris-Saclay, CNRS/IN2P3, F-91405 Orsay, France \label{ijclab}
\and
GANIL, CEA/DRF-CNRS/IN2P3, BP 55027, 14076 Caen cedex 5, France \label{GANIL}
\and
Normandie Université, ENSICAEN, UNICAEN, CNRS/IN2P3, LPC Caen, France \label{lpccaen}
}
\date{Received: date / Revised version: date}

\maketitle

\abstract{
Presently, $\gamma$-ray tracking in germanium segmented detectors is realised by applying two advanced, complex algorithms. While they have already triggered an intensive R\&D, they are still subject to further improvements. Making such algorithms effective, online in real time conditions and/or offline for deeper analysis, in data pipelines do require many additional software developments. This review paper gives an overview of the various bricks of software produced so far by the AGATA collaboration. It provides hints of what is foreseen for the next phases of the project up to its full configuration namely with 180 capsules in the array.
}
%
\setcounter{secnumdepth}{3} \setcounter{tocdepth}{2}


\section{Introduction}
The concept of gamma-ray tracking relies on two key algorithms. The Pulse Shape Analysis (PSA) processes digitised traces from electrically segmented germanium capsules to extract interaction points (3D positions, energies, times) at the crystal level while the tracking algorithm operates on the whole array to reconstruct, from the cloud of all interaction points, the trajectory of every single gamma ray emitted. An active R\&D has been and is still performed to enhance the performances of these two complex, advanced algorithms (see \cite{xref-tracking,xref-psa}). 

The main goal of this review is to highlight how such algorithms have been integrated in data pipelines and implemented for the processing. For online processing, the real time conditions put stringent constraints on the performances of the algorithms and on their integration. While these points are less crucial for offline re-processing, still the computing resources may be challenging as well as the amount of data to be handled. 

To run online, the two algorithms, but also many other processes around, should be optimised from the hardware and software point of view. As there might be different type or version of PSA/Tracking, the system should provide, as much as possible, an easy integration in data pipelines. On one hand, PSA is performed in local data pipelines : it means the processing of one particular germanium capsule requires only the data out of that capsule allowing thus parallel independent processes. On the other hand, the tracking algorithm is performed at global level because it is required to build time coincidences of all local levels including data produced by ancillary detectors. 

Since the coupling to many possible ancillaries and their acquisition system (see~\cite{GADEA_2011,Domingo_2012,CLEMENT2017,Dobon_2023} for more details), the content of the data pipe can be quite diverse. The data stream should be organised so that any process could/should run without having to deal with any input that is not required. 

The whole workflow could be described by Direct Acyclic Graphs (DAG). The nodes of the graph correspond to processes (called actors in the AGATA terminology) in charge of producing, processing and consuming the exchanged data, the edges of the graph representing the pipelines in which data are transported.

Online, the data processing chains have to be connected to the output of electronic cards : the system should be flexible enough to follow any evolution/modifications performed and to come~\cite{xref-eletronics}. PSA is a complex but also resource consuming algorithm. The current versions run on CPU based architectures. The real time processing can only be achieved by distributing tasks on a cluster \cite{xref-daq} : the processing logic of the DAG can be realised only by using advanced orchestration tools.  

Since data is read from files, the real time constraint is less demanding for the offline processing. So far single computers are used even if it would be more efficient to rely on clusters or computing clouds. Lighter orchestration tools (emulators in the AGATA terminology) able to manage part or even whole workflows can thus be used.

A rich ecosystem with many processes and services has to be set up to make sure raw data are well collected, processed, monitored, stored and distributed for re-processing.  AGATA is growing continuously with the ultimate goal to reach 4$\pi$ (180 crystals). It represents a regular investment over decades with from time to time periods of major upgrades/modifications.

Many software developments have been performed in the AGATA collaboration, and new are to be done, to make sure all the sub systems are properly configured, initialised, started and running. Monitoring the whole workflow at many different levels is mandatory. The system (network, disks, computers) is to be controlled and the integrity/quality of the data all along the processing chains has to be checked. Whenever a new ancillary detector is coupled to AGATA, new developments are needed in order to integrate or to build interfaces between the two systems. 

Specific software is also required to manage the produced data and meta data. It is also of paramount importance to deal with developments over decades. Maintenance, minor and major upgrades should be ensured. As well, deprecated libraries, technologies have to be replaced. Several collaborators in different laboratories/countries have to work on the same programs simultaneously. Depending on the objectives, adapted frameworks, approaches, multiple programming languages, might be used. History of modifications of the codes should be kept. The quality and reproducibility should be controlled, the building system for the compiled code should be reliable and as much as possible portable to various distributions in particular for re-processing.

In this review, all the sections are divided into two parts: the first one is dedicated to the technology as used to date and the second one to the modifications foreseen for the new AGATA Phase.

The first section is dedicated to the PSA/Tracking integration in data pipelines. The second section is devoted to the specific tools in charge of the workflow management while the third one provides additional,  non exhaustive, information related to the rich ecosystem that has to be set up for a fully operational workflow. The last section is about software quality and management.   

\section{PSA and Tracking : integration in data pipelines}
\label{sec2}
\subsection{Phase 1}
\subsubsection{Introduction}\label{sec2_1_1}
Running online PSA and tracking algorithms requires a computing power which can be achieved only using clusters. Since the processing is distributed, an advanced orchestration tool is mandatory.

From the beginning of the project, the AGATA collaboration has used the \textit{NARVAL} framework~\cite{Grave2005}. While written in \textit{ADA}, such framework allows to bind \textit{C/C++} codes through a predefined interface embedded in compiled and loadable shared libraries. Using this mechanism, most of the processing bricks (actors) developed by the collaboration so far have been written in \textit{C++} to be run on CPU based architectures. When required and possible, specific CPU optimisations, such as vectorisation or concurrency, have been implemented. 

All those bricks (and more) are grouped in a single project, called \textit{AGAPRO} for AGATA PROcessing~\cite{agapro}. It represents a library of actors used for online processing but also for offline re-processing. Sharing the same code in both environments is achieved by the use of the \textit{ADF} interface~\cite{adf}, which provides several levels of virtualisation. The concepts of this virtualisation are presented in section~\ref{sec2_1_2} while section~\ref{sec2_1_3} is dedicated to different components of the \textit{AGAPRO} package.

\subsubsection{Virtual interface to the data pipeline: ADF }
\label{sec2_1_2}

\textit{ADF} stands for AGATA Data Format/Frame/Flow. It is a standalone \textit{C++} library developed by the collaboration as an interface to plug-in new actors in AGATA workflows, to extract from the data pipeline the input required and to add the output produced. In object oriented programming, which is the case for \textit{C++}, inheritance provides a mechanism to build a hierarchy of classes : while the base classes allows the implementation of common features, the derived ones provide specialisation. \textit{ADF} contains base classes on top of which are built the different kind of actors (producer, filter, consumer) available in the \textit{AGAPRO} package. 

Concerning data, because AGATA is a long time project and coupled with many different ancillaries, it was crucial, since the beginning of the project, to avoid strong couplings between the processing actors and the underlying data pipelines. For that purpose, several layers of abstraction have been introduced. 

The first one concerns the way data are structured into the data streams. Data produced by any sub-system are separated from any others. They are uniquely identified by a key (a header) which precedes the payload in the data stream. In the AGATA terminology, a frame is composed of a key and a data part: the payload. Currently, the key part contains a type and a sub-type to identify the nature of the data and the producing sub-system, a high precision timestamp as delivered by the global distributed clock system (GTS)~\cite{Bellato2013}, the size of the data block and an additional sequence number. A set of frame (composite frame) can be grouped under a single, well identified key, allowing to build events, coincidences.

\begin{figure}[ht]
\centering\includegraphics[width=\columnwidth]{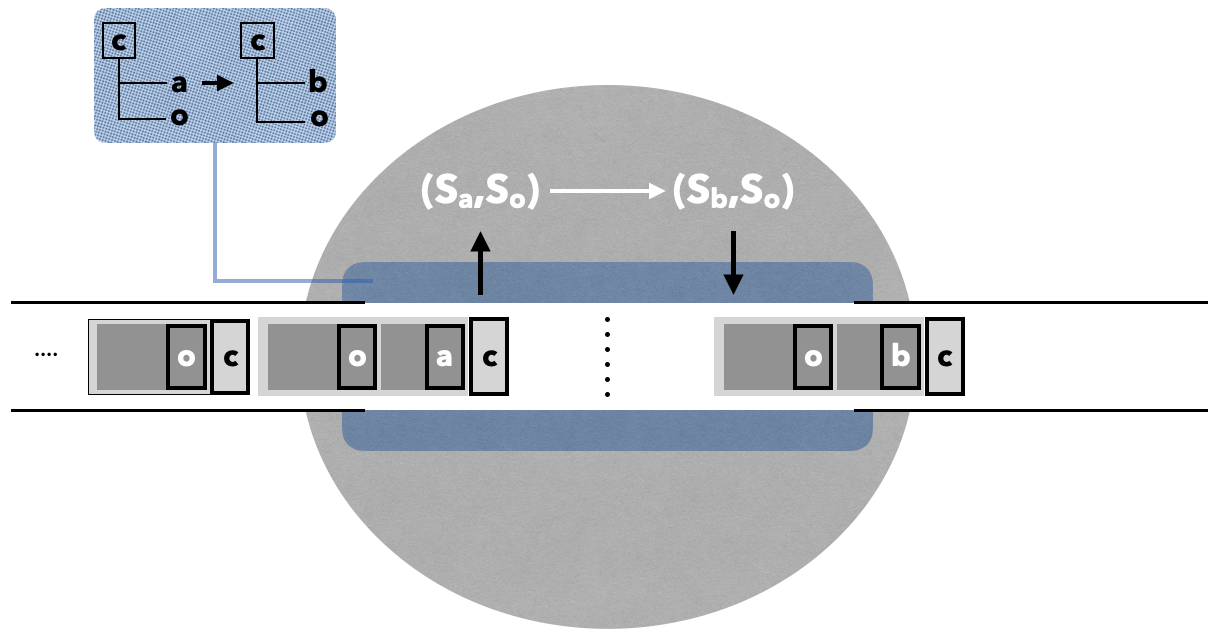}
\caption{Interface to the underlying data stream for a DAG node (grey circle). The top left (blue) box shows the trigger i.e. the full definition of what it expected as input and what is to be produced. At the initialisation phase, the trigger is registered to the \textit{ADF} data handler (blue box around the pipeline) which takes care of delivering to the process the relevant input information when available in high level in-memory structures $S_{a}$ and $S_{o}$ but also to stream out in a consistent way the ones produced.}
\label{fig:1}
\end{figure}

Knowing only the key, any process can then map the data stream without taking care of the payload (data part) content. This feature is the core of the second level of decoupling between the processing actor and the data pipeline and is realized through a trigger mechanism as illustrated in Figure~\ref{fig:1}.

In the illustration of Fig.~\ref{fig:1}, a node needs the inputs from two sub-systems $a$ and $o$ to compute the new $b$ output keeping in the stream the $o$ part. At the initialisation phase, the full trigger definition (as a sequence of keys) is registered to the \textit{ADF} data stream handler. Once data flow into the pipe, the handler checks whether of not the conditions are fulfilled, as it is the case for the first input frame of the drawing. It fills the corresponding high level structures ($S_{a}$ and $S_{o}$) and call the processing method of the actor giving the output structure ($S_{b}$). In case there are any computing issues, the record method of the \textit{ADF} handler takes care of adding consistently the output to the data stream. It should be mentioned that the \textit{ADF} handler does also deal with incoming data that are not specified by any trigger using a global policy resulting in either the consumption of the frame or its full copy on the output stream.

Serialisation/deserialisation of the data in the payload part are realised and managed by the \textit{ADF} library which provides the last level of decoupling : at the application layer, in-memory high level structures (such as digitised signals, hits, tracks) are provided to be used directly in users codes. Adding such a mechanism allows also to protect for any complications dues to architectures having different endianness. For a given general data interface, one can have different serialisation/deserialisation methods. This is handled by defining version numbers, for the key part as well as for the payload part of a frame. As version numbers are not part of the data stream, even if it could be added, the consistency should be managed at the global level. In particular for large payloads, as it is the case for digitised signals, one can define two versions, one with and one without compression, and switch from one to the other one depending on the resource (CPU, RAM, network) to be optimised.

\subsubsection{AGAPRO}
\label{sec2_1_3}

\begin{figure}[ht]
\centering\includegraphics[width=\columnwidth]{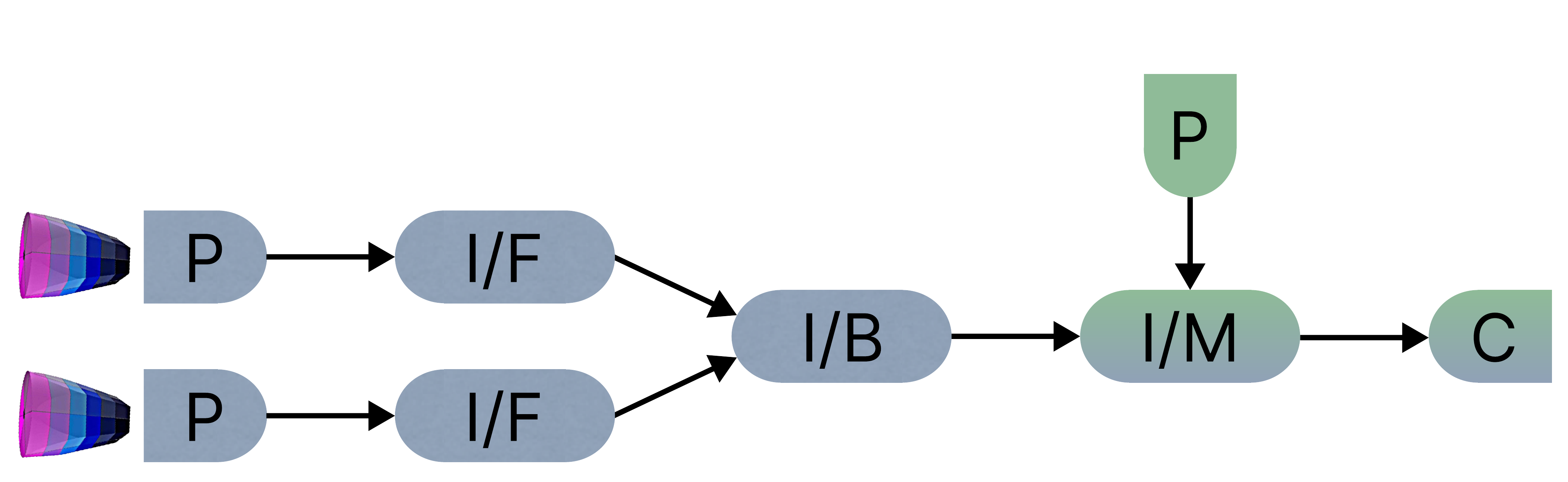}
\caption{Simplified graph for only two segmented germanium crystals (P), schematically drawn at the very left, of standard AGATA processing workflows. After a first processing (I/F nodes) the two independent lines feed a builder (I/B) in which time coincidence events are constructed. Then, it proceeds to through a merger (I/M) to build coincidences with the data coming from the ancillary detectors (P). At the end of the graph, the resulting stream is consumed (C).}
\label{fig:actor_pipeline}
\end{figure}

Depending of its position in the DAG describing the workflow, a node could be labeled as (see figure~\ref{fig:actor_pipeline}): 
\begin{itemize}
    \item Producer (P) in case it does not depend on any output from other nodes of the graph.
    \item Consumer (C) in case it does not produce any data to other nodes.
    \item Intermediary (I) if the node has input and output connections. In case of multiple inputs, the node could be associated to event builders (B, for AGATA events) or event mergers (M, for AGATA/ancillary events). With only one input and one output, it is also referred as a filter (F).   
\end{itemize}

Thanks to the abstraction levels introduced in section \ref{sec2_1_2}, nodes of the workflow can be internally updated, modified without affecting the processing of the other nodes: a rich library of actors can then be developed to manage different situations. They are all available and maintained within the AGAPRO package. 

Obviously, producers can be view as sources of raw data, coming from electronic readout, files, chain of files, compressed, using standard compression algorithms (such as \textit{gzip}), or by any protocol to exchange data over networks for instance TCP/IP. 

PSA and tracking are intermediaries actors, since they are connected to only two nodes, one being for input data, the second for output data. This kind of algorithm operates a deep transformation of the input data which can be consumed (i.e. it is not copied in the output stream) at the node level. Other filters are there to calibrate raw data (PrePSA actor) or to correct for unwanted effects (PostPSA actor for neutron damage corrections~\cite{Bruyneel2013}). Some processing do require to access the payload part to be run but others, such as time sorter, event builders, are agnostic and do rely only on the key part of the input data stream.

A basic consumer dumps the input stream as it comes (\textit{ADF} format) into files for storage and re-processing. The \textit{ADF} stream can also be transformed to build histograms or events in different formats (see at rear of section) adapted for specific analysis framework.  

Some nodes of the DAG do require more developments. For pure AGATA data, the first input node, the crystal producer, is strongly connected to very specific hardware consisting in PCIe cards developed by the collaboration~\cite{AKKOYUN201226,Egea_2017} and plugged in computers. In this node, the raw data can be internally dumped in files, in a format driven by the card, so that it can be read back later on for re-processing.

The PSA algorithm is computer-intensive and clearly requires to be optimised using all possible available technologies. Since the PSA actor is fed by bunches of independent events, it allows to parallelise the processing using multi-threading, each task being in charge of sub-bunches. Such feature has been implemented using first the \textit{boost}~\cite{boost} library and then the standard \textit{C++} thread library since the GANIL campaign. It corresponds to the period at which a substantial amount of the \textit{C++} AGATA software have been upgraded to fulfill \textit{C++11} standard. The very last stage of data processing should provide data that can be, as much as possible, usable for any user which is why a specific consumer (TreeBuilder) has been developed to save in \textit{ROOT} Trees~\cite{Brun_1997} the data of all subsystems including some complex ancillaries (see section \ref{sec4_1_1}).   

\subsection{Towards the $4\pi$ array}
\subsubsection{Introduction}
An intense R\&D is underway in order to improve the performances of the two key algorithms PSA and tracking. A promising avenue is to apply Machine Learning based technologies~\cite{xref-psa,xref-tracking}. This approach has a strong impact on the data processing model since, most of the time, it relies on heterogeneous hardware, such as Graphical Processing Units (GPU), but also on advanced third party libraries for instance \textit{TensorFlow}~\cite{tensorflow_2015} or \textit{Keras}~\cite{chollet_2015}. As well the available interfaces to deal with the resources are based on many different programming languages (\textit{Python}, \textit{Julia} etc ..) adding another level of complexity. 

The new Ethernet readout~\cite{xref-eletronics} is extremely demanding in computing resources in particular memory and network. With the evolution of the array to $4\pi$, High Performance Computing (HPC) infrastructure would be required to absorb efficiently the workflow. 

\subsubsection{ADF evolution}
The data flow structuring (see section \ref{sec2_1_2}) has proven to be well adapted to the different campaigns over more than a decade and thus can be used as a foundation for the futures phases. 

The \textit{ADF} library is currently based on the \textit{C++98} standard and has almost not been modified during the last GANIL campaign. Major modifications and upgrades of the code are to come in order to include new features provided by more recent \textit{C++} versions. Compression algorithms are to be implemented, eventually using native multi-threading, in particular for the frames containing the raw digitized signals, reducing the amount of data stored in memory and transported through the network by almost a factor 2.

A more general interface to the serialised data is foreseen using \textit{json} open standard~\cite{pezoa_2016} and lightweight data interchange format. Such new feature is mandatory because the AGATA collaboration progressively modifies its processing model to be more open data and open science compliant. This would probably simplify the binding to other languages. In particular, \textit{python} interfaces to \textit{ADF} data seems highly required to deal with Machine Learning based approaches. Possibly, the \textit{ADF} library could evolve and be based on recent technologies such as \textit{Protocol Buffer}~\cite{protobuf} (or \textit{FlatBuffer}~\cite{flatbuffers}), an open source cross-platform data format used to serialise data. 

\subsubsection{AGAPRO evolution}
As part of a global ecosystem, all the bricks of software available in the \textit{AGAPRO} package are likely to evolve. As for \textit{ADF}, the current \textit{AGAPRO} package is to be modified to take advantages of modern coding techniques and tools. In particular, when possible, 'home made' code should be replace by highly optimised library code, for instance standard \textit{C++} libraries. To offer more possibilities it is planned to include the \textit{boost} library (which very often contains the future standard versions of \textit{C++}) as main dependency.

As it has been done for PSA, concurrent programming techniques are to be generalised to try and optimise various aspects of the data processing chain. New bricks are to be developed to deal with the new electronics readout~\cite{xref-eletronics}, the new monitoring system (see section~\ref{sec_glob_monitoring}) but also to deal with heterogeneous hardware (GPU, FPGA) likely to be used for PSA or tracking algorithms. The new \textit{C++} part of the package is to be developed with the latest standards keeping as a guideline that code has to be robust and optimised. Because of more heterogeneous hardware, the use of non \textit{C++} code is likely to grow in the future. To better manage the increasing number of dependencies, languages and tools, as it has been done already for different aspects (see section~\ref{sec5}), it is planned to make more extensive use of virtualisation technologies, lightweight containers, such as \textit{docker}~\cite{Merkel_2014} or \textit{singularity}~\cite{Kurtzer_2021}. 

\begin{figure}[ht]
\centering\includegraphics[width=\columnwidth]{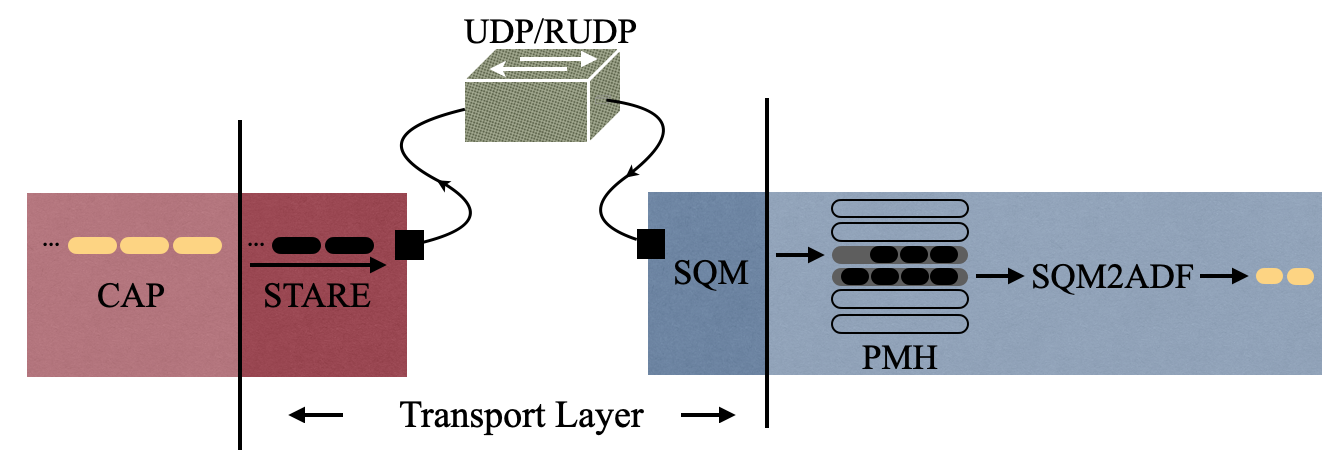}
\caption{Main ingredients of data pipeline of the new electronics readout. ADF frames are produced and formatted by the pre-processing card CAP, the transport over the network, using the UDP protocol, being managed by the STARE board. A daemon collects the UDP packets to reconstruct in RAM the data stream as produced by the CAP board. The SQM2ADF process consumes then the events in the RAM and is then the first node of the processing graph.}
\label{fig:V2Datapipeline}
\end{figure}

The new upcoming electronics board has already triggered many software developments requiring cautious optimisations. The main ingredients of the associated data pipeline are represented in Figure~\ref{fig:V2Datapipeline} (see~\cite{xref-eletronics} for more details on the electronics). On the PACE board (left part of the figure), the STARE mezzanine is in charge of sending data, already formatted using the \textit{ADF} format, to its counter part on computer, SQM (for STARE Queue Manager), through Ethernet using the UDP protocol~\cite{Postel_1980} to ensure very high transfer rates up to the maximum bandwidth of the link. Contrary to TCP/IP~\cite{Postel_1981}, the UDP/IP protocol do not guarantee a packet is delivered, neither the delivery order. On top of UDP, a reliable UDP (RUDP)~\cite{Bova_1999} protocol is implemented on both sides to strengthen the data transport: this requires to add (done at the STARE level) an UDP header to any piece of data pushed in the data stream by the Control And Processing board (CAP). It should be noted that the standard size of an \textit{ADF} frame encoding the full signals out of one AGATA crystal is 8088 bytes which fits into the maximum size of an Ethernet jumbo frame, avoiding thus any drops in performances possibly due to fragmentation into multiple Ethernet packets. 

As part of the transport layer, SQM should reconstruct the data stream ordered as it is given to STARE by CAP. Currently, SQM fills large buffers in computer RAM. Once one of them is full, \textit{DCOD} (see section~\ref{sec3}) starts its process by calling a dedicated actor (SQM2ADF) which is then a first node of the global workflow. 

Because of the extremely high rate performances of the new AGATA board, both SQM and SQM2ADF should be optimised. It has been decided to reduce as much as possible the number of actions handled by SQM. Its main, but nevertheless complex, goal is to order in real time the data stream delivered by STARE to recover CAP input. While the UDP header should or might be removed before pushing to RAM, it has been decided to keep it as redundancy to simplify the SQM2ADF tasks. To handle the highest rates up to almost the hardware limit of $10Gbits/s$, SQM should be run cautiously. The CPUs used should not be the ones dealing with interruptions to the operating system. As well the UDP packets should be pushed to the RAM bounded with the CPU in charge of SQM. 

SQM2ADF has the charge to consume as fast as possible the data stored in RAM by SQM so that it can be sent to the next processing actor of the chain. Several threads can be run in parallel to try to speed up the task. Depending on the kind of data, appropriate actions are taken with some checks of the consistency of the \textit{ADF} frame. Another important feature is the possibility at this level to compress the \textit{ADF} frame in order to reduce the imprint in memory but also to decrease the amount of data transferred over the network. Thanks to a very specific compression algorithm~\cite{Buerger_private}, up to a factor two in size could be gained with a quite low CPU overhead. Preliminary studies, using current DAQ computers, have shown that three to four threads are needed to handle compression at rates up to $50kHz$ which is the highest established limit. It illustrates how concurrent programming techniques are going to be the default scheme for any future developments in order to add flexibility and to ensure optimal processing in various situations.  

Up to now, the individual bricks of the new detector level data pipeline shown in Figure~\ref{fig:V2Datapipeline} have been mostly developed. It remains to debug them, to make them robust for production and to integrate all of them properly in real-time data pipelines. 

\section{Workflow Management}
\label{sec3}
\subsection{Introduction}
As already mentioned, \textit{NARVAL} and its more recent evolution \textit{DCOD}, is the orchestration tool selected by the collaboration and it has proven, for more than a decade, to be a solid foundation, flexible enough for the various campaigns performed so far. With the inclusion of heterogeneous hardware, the evolution toward HPC and all the developments foreseen for the $4\pi$ version of the array, it is to be enriched. This is also an opportunity to built more advanced offline reprocessing solutions keeping always the same approach,  which consists of building bricks shareable between both online-offline environments. 

\subsection{Phase 1}
\subsubsection{Online}
Online, nodes of the workflow graphs have to be connected to the electronics readout and external data acquisition systems, such as \textit{MBS}~\cite{Essel_1999} at GSI, \textit{NARVAL}~\cite{Grave2005} at GANIL and \textit{xDAQ}~\cite{Brigljevic_2003} at LNL. Before being executing, one has to be sure that the whole workflow is consistently built. A dedicated tool, the \textit{Topology Manager}, has been developed by the collaboration for that purpose. It is written in \textit{ADA}, based on a \textit{PostgreSQL} database~\cite{postgre} to describe devices and the way they are connected. A web embedded interface allows to fully configure the processing graph and generates the configuration files using most of the time the \textit{XML} language~\cite{XML}. 

Based on these files, the \textit{DCOD} orchestration tool could be initialised. PMH, for Posix Memory Handler, is an important evolution, realised within the collaboration, from \textit{NARVAL} to \textit{DCOD}. It emulates an efficient, distributed shared memory system using the memory of the nodes of the cluster and ensures the data workflow is consistent with the chosen data policy. It prefigures other open-source tools such as \textit{Memcached}~\cite{memcached}, \textit{Redis}~\cite{redis}, \textit{Dhmem}~\cite{dhem}, which ensure similar virtual shared memory but for containerised workflows. CTL, for Common Transport Layer, is another component of \textit{DCOD} allowing to efficiently transport in memory buffers from one cluster node to another one so that it can be processed. 

Through a client running in consoles, one can control the workflow execution. However, for non expert users and thanks to embedded web services, it has been made possible to build Graphical User Interfaces (Run Control) using various environments such as for the first Legnaro and GSI campaign \textit{C++/Qt}, \textit{Java} for the GANIL one and \textit{python/Flask} for the current one. 

\subsubsection{Offline}
The development of new actors, often written in \textit{C/C++} (see section \ref{sec2}), requires executables to run in debugger tools. To allow this, \textit{C/C++} programs or libraries have been written to emulate simple orchestration tools. They are called emulators in the collaboration and they can be found into the \textit{AGAPRO} package (\textit{femul}, for flat emulator~\cite{dino_private}) or in the \textit{GammaWare} package~\cite{gammaware}. Another possibility is to use \textit{Narval Standalone}, a simplified \textit{DCOD} environment. Such tools can run only on single computer. Even if some of them can deal with full workflows, they are seldom used as such for evident reasons of performances. Multi-threading could be activated to accelerate the execution. It might be enough for relatively small workflow. With more that few tens of capsules, the limits of such approach is already outreached. Thus most of the time, the full workflow is decomposed in sub-parts processed independently and intermediate output data being stored in \textit{ADF} files. 

\subsection{Towards the $4\pi$ array}

Machine Learning based algorithms are being studied in order to improve PSA and tracking performances~\cite{xref-psa,xref-tracking}. In particular, neural networks have the potential to speed up greatly the inference time for PSA. However, such algorithms likely require to add heterogeneous hardwares (GPU, hybrid CPU-FPGA chips) in the global workflow turning the DAQ box into a full HPC farm. Another drawback is the advanced libraries, such as \textit{TensorFlow}~\cite{tensorflow_2015}, \textit{Keras}~\cite{chollet_2015} or \textit{PyTorch}~\cite{NEURIPS2019_9015}, on which the neural networks are built. The current AGATA workflow should be adapted to deal with these new constraints. A first step in that direction has already been reached in a similar environment during the AGATA-NEDA-DIAMANT campaign at GANIL~\cite{Fabian_2021}. More steps are mandatory and, in particular, one should be sure the different hardware are used in the most efficient way. 

This point, together with the new electronics boards with its Ethernet readout, has motivated new developments to add more flexibility in the global workflow. A dynamic load balancing would allow for instance to allocate more resources in case the processing is not fast enough. As well, events can be dispatched to different HPC nodes depending on their topology: advanced PSA, on GPU, could be used for multiple hits inside the crystal while a more standard, CPU based, can be run for single hits. 

\begin{figure}[ht]
\centering\includegraphics[width=\columnwidth]{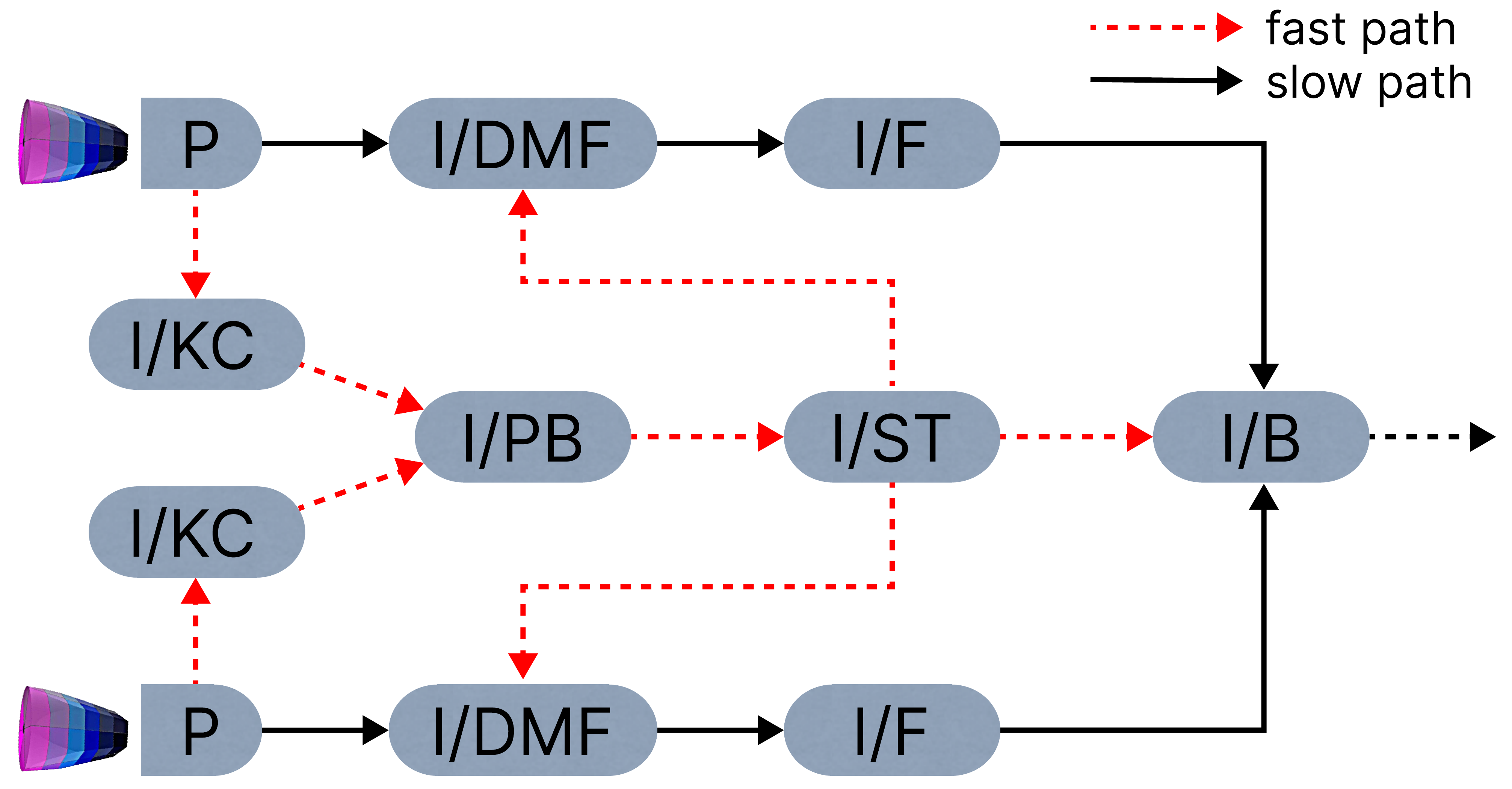}
\caption{Schematic graph to implement a global software trigger for data produced by two germanium crystals. For each crystal, the data stream is separated to follow two paths, one labeled as `slow' (black straight lines), similar to the one shown in figure~\ref{fig:actor_pipeline}, and a second path (red dashed lines) referenced as `fast'. On the `slow' path, the PSA algorithm is applied which allows the `fast' one to be executed in advance. For this, the input `fast' stream is first filtered (I/KC) to keep only a fraction of the ADF frame, for instance the key part (KC meaning Key Collector). All the `fast' streams are merged (I/PB, PB being for Pattern Builder) to build coincidences which are sent to the Software Trigger (I/ST) for validation. Only validated data are released at the Data Manager Filter node (I/DMF) to be processed on the `slow' path.}
\label{fig:SoftwareTrigger}
\end{figure}

To further optimise the PSA processing load, a software trigger, at the global level, is to be used with the goal to process only the most interesting events. From the producer nodes, two paths of data are defined. One (`fast path') with only a small fraction of the ADF frames (for instance the key part) directly send to a global pattern builder, building time coincidences, and then to a software trigger validating or not the events that are processed trough the standard pipeline (`slow path'). For online processing, the \textit{DCOD} workflow manager has been designed to handle all these possibilities quite naturally, and a first version of a software trigger has been tested. 

With such a growing complexity of the workflow, and since the offline tool currently used has already reached its limits, it is important to provide a new, scalable, environment. As more and more applications are containerised within the collaboration, docker, for single computer, and docker swarm for clusters, seem a natural solution currently under investigation. Possibly this solution could be adapted also for the facilities offered by the European Open Science Cloud (EOSC)~\cite{EOSC}. 

Whatever the official selected solution for offline re-processing, to be compliant with Open Sciences directives, the workflow scheme should be easily exportable into different computing environments. This could be done by following standard solutions such as Common Workflow Language\cite{commonwl}. To achieve this goal, the AGATA collaboration has already developed the appropriate application: the Topology Manager, able to build consistent workflow for online processing. What remains to be done is to export it in different suitable formats.

\section{Other bricks of the ecosystem}
\label{sec4} 
\subsection{Introduction}
One one hand, the global workflow is connected to data producers, electronic boards and ancillaries data acquisition systems for online, mainly files for offline. On the other hand it ends with data consumers, most of the time dumping data into files in various format. Thus many software bricks are also required to make sure the electronic boards are providing good data but also to deal with the cycle of life of all the data and meta data once produced. 

All the processes involved should be controlled with a monitoring system able to ensure data integrity and quality. With the important amount of data produced, data mining tools are also to be developed to fully exploit the potential of the array. 

\subsection{Phase 1}\label{sec4_1} 
\subsubsection{Applications associated to nodes}
\label{sec4_1_1}
Many software tools have been developed and deployed to deal with the first generations of electronics boards requiring to plug PCI express cards into computing nodes. Since the new board, with its Ethernet readout and IPBus Slow Control, should make them progressively obsolete, they are not covered in this paper. Of course, it remains important to maintain them and make sure they can be used as long as they are needed. 

Most of the time, the coupling with local acquisition systems has been performed thanks to dedicated producers which were in fact linked, though the TCP/IP protocol, to external (from the point of view of the global processing workflow) processes delivering stream of raw ancillary detector data encapsulated with an ADF header. Most of the workflow is agnostic to the real content/format of the ancillary frame.

The full procedure to calibrate properly AGATA has also driven the development (or adaptation) of many tools/applications. Energy calibration, time alignment, cross talk and neutron damage corrections are well established and documented. The procedures to be applied and the appropriate codes are part of the \textit{AGAPRO} package. Graphical user interface, such as tkt~\cite{dino_private} or cubix (included in the GammaWare package)~\cite{gammaware}, are also extremely used to help for such complex tasks.  

The global workflow ends with a single consumer which converts the events from the \textit{ADF} format into entries of the extremely popular, widely used and well supported, \textit{ROOT/TTree} format~\cite{Brun_1997}. Conversion of the raw ancillary data, applying calibration and even advanced reconstructions for the most complex array such as the GSI-FRS~\cite{GEISSEL_2003,WINKLER_2008}, PRISMA~\cite{Stefanini_2002,Szilner_2007}, VAMOS~\cite{Rejmund2011,Kim2017} or MUGAST~\cite{ASSIE2021}, is performed at this level by linking when necessary external compiled libraries~\cite{vamoslib,prismalib,Matta2016}: it results for the last stage analysis, chains of \textit{ROOT/TTree} composed of several sub-branches associated to the different sub-systems. It should be noted that many collaborations have based their last stage analysis on the \textit{ROOT} format which has greatly simplified their integration into the AGATA data flow.   

\subsubsection{Global monitoring}
To ensure data integrity and quality, online monitoring has evolved during the different campaigns. At the very beginning, it relies on intrinsic functionalities of \textit{DCOD} able to deliver to clients, through embedded TCP/IP server, a copy of the output buffer as produced by the various processing nodes. Parts of the GammaWare package~\cite{gammaware}, were dedicated to collect those buffers which where then processed using the same code than for offline re-processing. One drawback of this approach in the sampling of the data buffers performed at the \textit{NARVAL} level to avoid slowing down too much the processing of the nodes. Another system has been developed consisting in building directly all the required spectra inside the processing nodes. Saved regularly in files, external applications were able to read them back to control the workflow. Such way to proceed allows control spectra with the full statistics to be built. The main inconvenience is the time required to save all the spectra on disk : this way is certainly not scalable up to $4\pi$. 

\begin{figure}[ht]
\centering\includegraphics[width=\columnwidth]{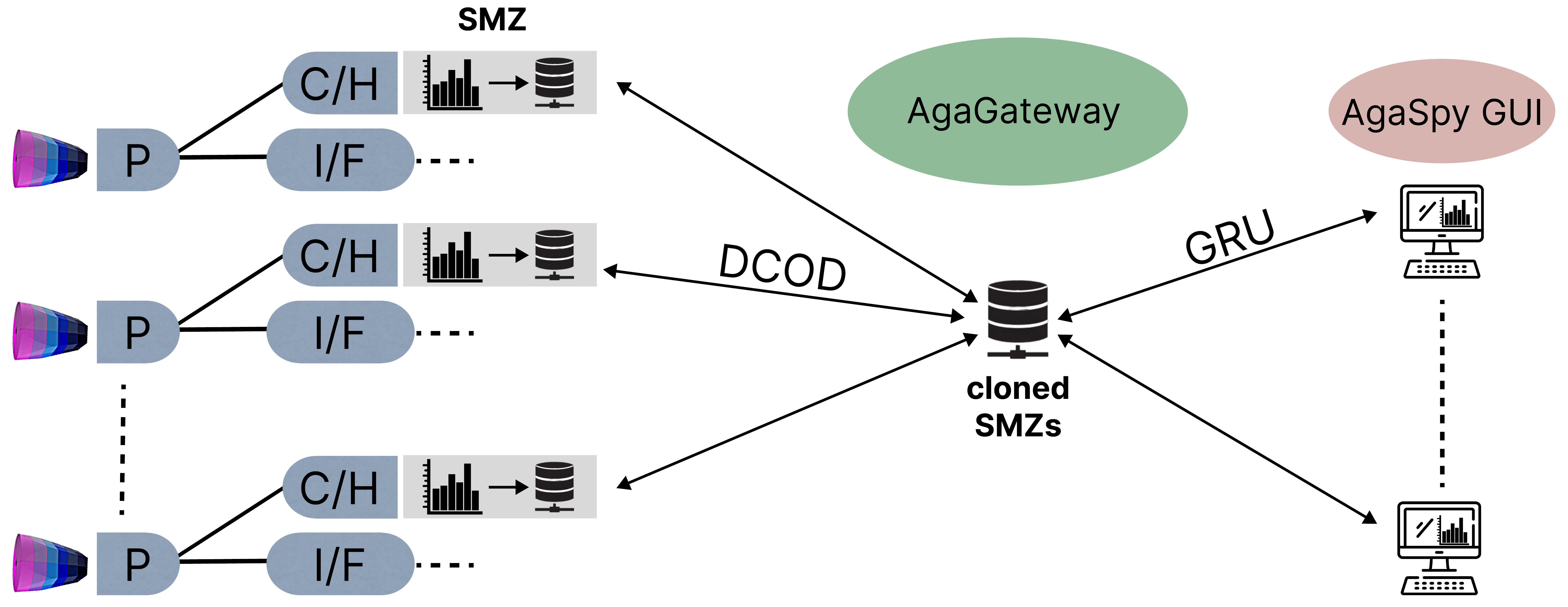}
\caption{Schematic representation of the AgaSpy monitoring.}
\label{fig:agaspy}
\end{figure}

\begin{figure}[ht]
\centering\includegraphics[width=\columnwidth]{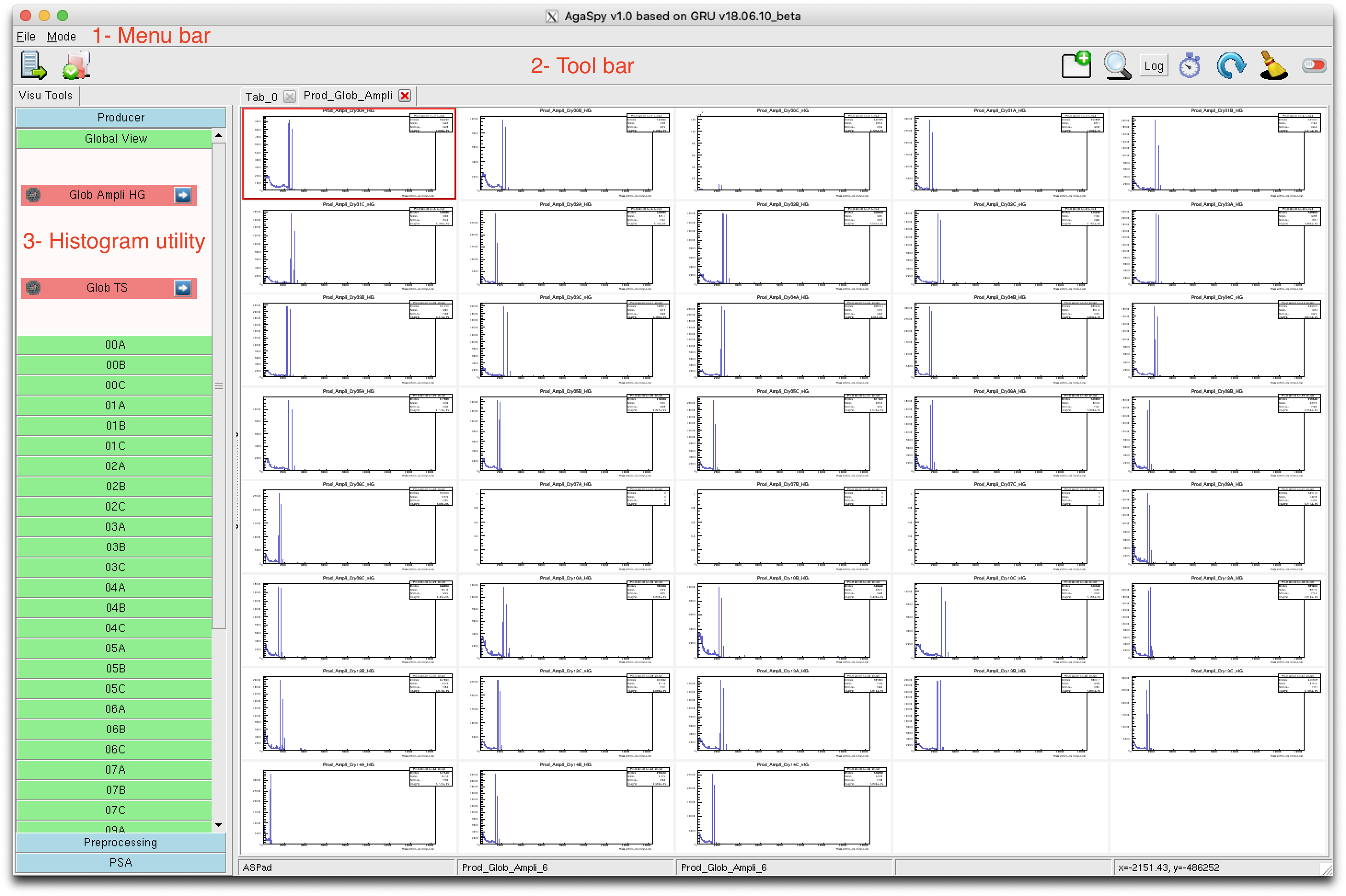}
\caption{AgaSpy graphical user interface.}
\label{fig:agaspy_gui}
\end{figure}

The current method, the AgaSpy tool~\cite{agaspy}, has been developed during the GANIL campaign to overcome this difficulty by using the distributed shared memory facility provided by \textit{DCOD}. As represented on Figure~\ref{fig:agaspy}, the output of each node of the processing graph is duplicated by \textit{DCOD}, the copy being transported to dedicated histogramer consumers (C/H) in charge of building the appropriate spectra directly in the distributed shared memory zone (SMZ). An additional central process named AgaGateway collects all the spectra produced and make them available to any other computer on the network through a GRU Spectra server~\cite{GRU}. A dedicated GUI has been developed to easily monitor the spectra for the different detectors and nodes (Figure~\ref{fig:agaspy_gui}).

\subsubsection{Data and meta data management}
Data are produced, and thus can be saved, all along the workflow~\cite{xref-daq}. A great deal of meta data is thus required to fully master the data content. The configuration files of any node (calibration coefficients, algorithm parameters, etc ...) represent the minimal information to be handled and saved. This is so far performed by a single \textit{python} script used as a global database. The parameter of the script are directly modified in the code to produce a new set of configuration files. For each run, the script, the meta-data files as well as the data produced are saved consistently using sub-directory structures.   

Once an experiment is over, the full dataset (top directory) is moved to GRID data centers (CNAF Bologne and CCIN2P3) using \textit{python} scripts based on the \textit{gfal} (Grid File Access Library) python library~\cite{gfal}. As well, datasets can be downloaded by similar scripts to be re processed by any collaborator part of the AGATA Virtual Organisation.   

\subsection{Towards the $4\pi$ array}

\subsubsection{Applications associated to nodes}
The ecosystem of programs around any nodes of the processing graph is likely to change in case new PSA or tracking algorithms are available~\cite{xref-psa,xref-tracking}. Machine Learning based approaches, especially for supervised methods, are particularly challenging concerning the calibration phase (see also Data and Meta data management). Indeed, the training phase should be mastered to avoid any over-fitting, to optimise all the hyper parameters and to ensure reproducibility. \textit{Jupyter notebook}~\cite{Jupyter_Notebooks}, or similar, are suitable environments to be used to develop the code to manage this part.      
Regarding producers, connections to external acquisition systems should be quite easily handled as it has been done so far, with the tools already developed, and details have already been given regarding the new electronic foreseen.

Possibly, new consumers are to be written to save data in different formats suitable which would allow to use other last stage data analysis environments. Hierarchical Data Format (\textit{HDF}~\cite{hdfgroup}) can be considered as well as dataframes based files that can be easily processed using modern, users friendly and highly efficient frameworks such as \textit{Pandas}~\cite{pandas}, \textit{R Studio}~\cite{RStudio}, \textit{Spark}~\cite{spark} or \textit{Hadoop}~\cite{hadoop}.
Whatever the future last stage data analysis frameworks, with the growing of the array, tools to quickly handled high-fold $\gamma$-ray coincidence data, set up for the previous generation of $\gamma$-ray spectrometers~\cite{Cromaz_2015,STEZOWSKI_1999} should be adapted to modern computer architectures. AGATA and GRETA have just opened the path to new and innovative analysis methods (with their implementation) to fully exploit the potential of $\gamma$-ray tracking.    

\subsubsection{Global monitoring} 
\label{sec_glob_monitoring}

With the increasing number of germanium capsules in the array, the expected more complex algorithms possibly requiring heterogeneous hardware, the additional stress added to the network (cards in computer, switches) and RAM by the new electronic board, new tools are mandatory to properly monitor and control the workflow. This is for the collaboration also an opportunity to unify the online and offline monitoring. Spectra are still to be built at different levels in shared distributed memory as it is currently done and this facility is to be extended for offline re processing. 

The new additional direction followed is to include time series databases, such as \textit{InfluxDB}~\cite{InfluxDB}. This phase has already started for some sub-systems and is to be generalised. The data collected are to be summarised in dedicated dashboards using the \textit{Grafana}~\cite{grafana} or similar tools. This may represent a huge amount of data to be visualised and checked : it is thus foreseen to deploy modern search and analytic engines such as \textit{Elastic Search}~\cite{elasticsearch}. Since any data, parameters, in the database are timestamped, LSTM neural networks~\cite{HochSchm97}, or associated, may also be set up to try and predict issues before they can corrupt the data stream.    

\subsubsection{Data and meta data management}
With the new phase, AGATA has started a process to make its data FAIR~\cite{FAIR2016} i.e. Findable, Accessible, Interoperable and Reusable : this is a first step to open data and open science perspectives. Progressively the workflow is to be adapted/modified. In particular much more meta-data are to be managed using computer readable standard formats as \textit{XML} or \textit{JSON}. As well, the meta-data are to be saved in appropriate (relational, document or even graph oriented) databases for long time preservation. 

A common and widely spread practice in High Energy Physics to organise a great deal of meta-data (configuration, calibration, algorithm parameters) is by using Intervals Of Validity IOV. An IOV provide information on a time window during which a given set of parameters (named as payload) are pertinent to the associated data set. It can be some run, sub-run, identification numbers or universal timestamps as the one delivered by the GTS system~\cite{Bellato2013}. Relational databases provide good ways to implement such a design. Without any doubts, AGATA would greatly benefit of this approach to handle more precisely, efficiently, the workflow at the cost probably of many new software developments. 

In parallel, data catalogs, using application such as \textit{Rucio}~\cite{rucio}, of the datasets produced should be deployed on accordance with the OpenNP initiative funded through the EURO-LABS European project~\cite{eurolabs}.

\section{Software management, quality and distribution}
\label{sec5}
\subsection{Introduction}
Projects involving many collaborators are likely to change and since many software developments are required, it has been important to have suitable environments to maintain, upgrade, modify, ensure quality, reproducibility and portability of the various bricks of code.

Following the evolution of the standard languages (\textit{ADA95} to \textit{ADA2012}, \textit{python2} to \textit{python3}, \textit{C++98} to \textit{C++11} and soon \textit{C++17} or \textit{c++20}) and associated libraries is also a mandatory task. 

For its phase 2, AGATA is following the path to open science. It requires some evolution of the current models, practices, to achieve such a goal.  

\subsection{Phase 1}
Free, open source version control systems have been used since the beginning for most of the project developed within the collaboration. Tools like \textit{cvs}, \textit{svn}, \textit{hg} or more recently \textit{git}~\cite{chacon2014pro} have played a central role in software management keeping track of modifications of code for more than ten years. 

\begin{figure}[ht]
\centering\includegraphics[width=\columnwidth]{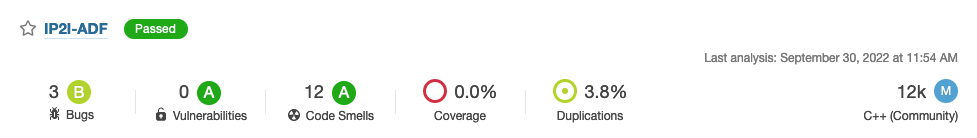}
\caption{SonarQube dashboard representing the global health assessment of the \textit{ADF} package.}
\label{fig:sonarcube1}
\end{figure}

\begin{figure}[ht]
\centering\includegraphics[width=\columnwidth]{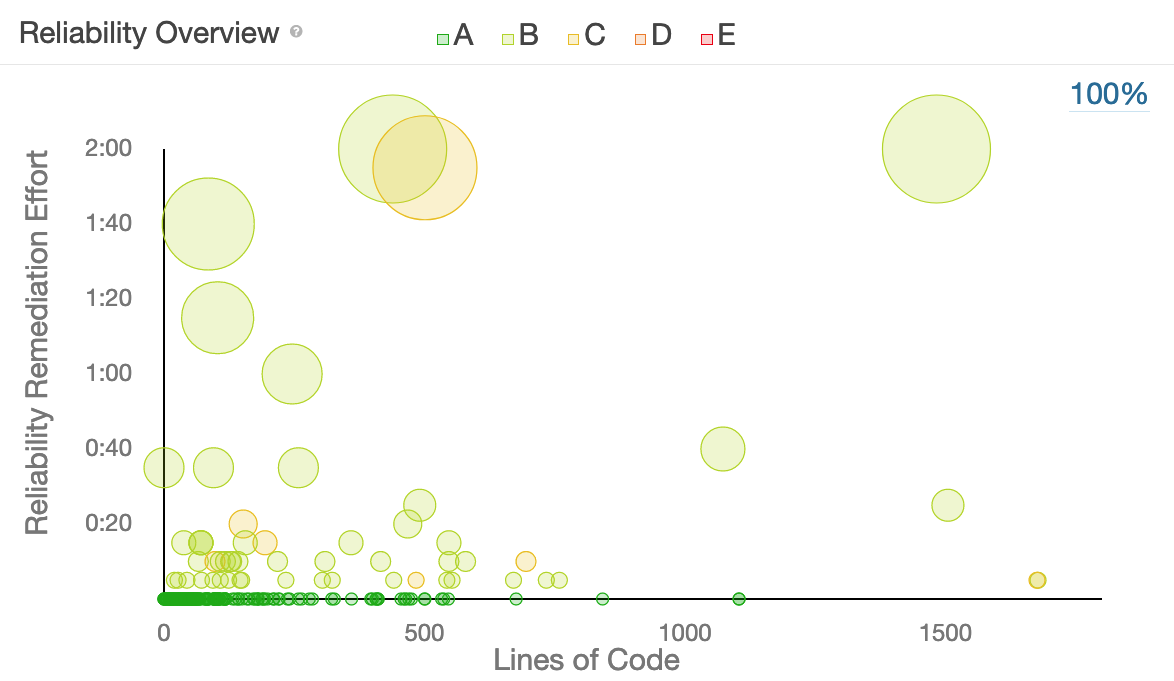}
\caption{SonarQube dashboard representing the detailed view to track particular pieces of code in the various files of the \textit{AGAPRO} package.}
\label{fig:sonarcube2}
\end{figure}

Since compiled codes are mostly used, compilers are the first tools to check their quality. In particular, the \textit{ADA} language and its compiler being extremely rigorous, it ensures less error prone binaries. A different strategy has been deployed for \textit{C/C++} codes:  the more recent environments, \textit{Github}~\cite{GitHub} or \textit{Gitlab}~\cite{GitLab}, allow to set up automatic validation pipelines including continuous integration, deployment and more. So far, for the \textit{AGAPRO} / \textit{ADF} and some others packages, and thanks to docker containerisation, modifications are pushed to the central repository only if the project compile without errors in standard LINUX environments. At the same time, code quality metrics are also realised using, in our case, the combination of two tools, \textit{Cppcheck}~\cite{cppcheck} and \textit{SonarQube}~\cite{sonarqube}, Figures~\ref{fig:sonarcube1} and~\ref{fig:sonarcube2} represent snapshots of the final dashboards obtained on SonarQube respectively for the \textit{ADF} and \textit{AGAPRO} packages. The first one provides global health assessment while the second one provides a more detailed view to track particular pieces of code in the various files. 

In some cases, the running codes have been profiled using tools such as \textit{valgrind}~\cite{valgrind} in order to look for bugs, memory leaks, bottlenecks but also to optimise the execution for particular architectures. This is explained in details for the PSA code in \cite{CALORE_2013}. Similarly, the Orsay Forward Tacking~\cite{Lopez-Martens2004} has been profiled to identify and optimise the implementation of complex formula.  

The online processing requires to manage a cluster likely to evolve and be composed of different computers and operating systems. To deal with this constrain, a specific application \textit{PEM}, for Project Environment Marketplace, has been developed within the collaboration to provide a coherent environment (set of pre-compiled tools) for multiple LINUX operating systems and different architectures.

Thanks to version control systems, the codes produced have been distributed worldwide quite easily. Advanced building systems, based on \textit{Makefile} at the very beginning and quickly \textit{CMake}~\cite{CMake} (including its modern version recently) have been maintained and proved to be solid enough over the years. However, with the recent advances in virtualisation, virtual machines and containerised applications are more and more used and should be the main way to distribute in the future.     

\subsection{Towards the $4\pi$ array}
The AGATA collaboration has already applied some of the Development and Operations paradigm (DevOps) practices~\cite{Jabbari_2016}. The goal of the Phase2 is to strongly enhance these practices, to have a more global and coherent approach in order to shorten as much as possible the development cycle and to provide to the user high software quality. With open science as a long term objective, the collaboration should join initiatives as proposed for instance by the Software Heritage~\cite{softwareheritage}.

A more systematic use of services such as \textit{Github} and/or \textit{Gitlab} are foreseen. While for the moment continuous integration processes have been set, we would like to extend them to cover also continuous delivery and deployment: unitary tests are to be added if necessary, and key codes are to be run automatically to check for their reliability on reference data sets. 

With more heterogeneous hardware in the workflow, it is also required to really optimise their use. As already mentioned in~\cite{CALORE_2013}, vectorization, in particular for PSA, can be a complex operation to implement and master. To tackle this problem, studies are ongoing in the collaboration using dedicated tools like \textit{MAQAO} (Modular Assembly Quality Analyser and Optimizer)~\cite{maqao} to try and accelerate the evaluation of the Figure of Merit for the grid search algorithm of the PSA. Another direction currently under investigations is to make use of low precision arithmetic in order to reduce the in-memory imprint, to minimise the energy consumption and to increase the processing speed. Of course, such approach requires to make sure it does not degrade the accuracy and stability of the results more that what is acceptable from the physics point of view: tools like \textit{CADNA} (Control of Accuracy and Debugging for Numerical Applications)~\cite{cadna} are under evaluation for that in the AGATA collaboration. 

As already mentioned, several applications produced are already distributed widely using containers (\textit{docker} or \textit{singularity}). This practice should be extended in the collaboration with the objective to deliver to users applications optimized for specific (HTC, HPC or clouds) environments. 

\section{Conclusions}
For more than ten years, AGATA has produced data and this, together with the major upgrades realised, has driven many software developments. With such long time scales, software development life cycle should be managed conscientiously. As described in this paper, from the deep integration of the key AGATA algorithms up to the most global levels, it has been realised using advanced, mostly open source, standards and libraries. While the general data processing model has evolved, it has not changed strongly in its foundations (CPU based, HTC cluster). With the coming of the new Ethernet readout, the advances in Machine Learning approaches and the uses of heterogeneous accelerators (such as GPU or FPGA), the AGATA collaboration has to solve these new challenges in software developments on the path that will bring it to a full open science horizon. 

\section*{Acknowledgments}
The authors would like to thank the whole AGATA collaboration. The production of all the essential bricks of software, their maintenance and optimisation is the result of a constant and tremendous amount of hard work involving a huge number of people in many European laboratories. Particular thanks go also to the skilled engineering and technical staff at the various host facilities for taking the additional charge of running the system during the physics campaigns.        
Part of this project has received financial support from the CNRS through the MITI interdisciplinary programs.

\bibliography{bibliography}
\bibliographystyle{myapsrev4-1}

\end{document}